\documentstyle[11pt]{article}

\newcommand{\SI}{\Sigma}

\newcommand{\xn}{x_{n}}

\newcommand{\xm}{x_{m}}

\newcommand{\kim}{ k_{1}^{\mu}}                                      
\newcommand{\kom}{ k_{0}^{\mu}}                                      

\newcommand{\yn}{ Y_{n}}                                             
\newcommand{\ym}{ Y_{m}} 
\newcommand{\kn}{ k_{n}}

\newcommand{\km}{ k_{m}}

\newcommand{\yim}{ Y_{1}^{\mu}}                                      
\newcommand{\yin}{ Y_{1}^{\nu}}                                      
\newcommand{\kin}{ k_{1}^{\nu}}                                      
\newcommand{\kon}{ k_{0}^{\nu}}                                      
\newcommand{\ktm}{ k_{2}^{\mu}} 
\newcommand{\ktn}{ k_{2}^{\nu}}                                      
\newcommand{\ytm}{ Y_{2}^{\mu}}

\newcommand{\lpp} {e^{i \int _{c} \alpha (s) 
k(s) \partial _{z} X(z+s) ds +ik_{0}X(z)}}

\newcommand{\gvk}{ e^{i\sum _{n \ge 0 }k_{n}\tilde Y_{n}(z)}}
\newcommand{\gvkl}{ e^{i\sum _{n \ge 0 }k_{n} Y_{n}(z)}}

\newcommand{\p}{\partial}                                           
                                           
\newcommand{\pp}{\partial ^{2}}

\newcommand{\eps}{ \epsilon}                                        
\newcommand{\al}{\alpha }                                             
\newcommand{\aln}{\alpha _{n}} 
\newcommand{\tY}{\tilde Y}

\newcommand{\la}{\mbox{$ \lambda $}} 
\newcommand{\be}{\begin{equation}}
\newcommand{\br}{\begin{eqnarray}}
\newcommand{\ee}{\end{equation}} 
\newcommand{\er}{\end{eqnarray}} 
\newcommand{\eln}{\mbox {$ e^{\sum _{n}\lambda _{-n}L_{+n}}$}}

\newcommand{\ppp} { \partial ^{3}}

\begin{document}
\renewcommand{\theequation}{\thesubsection.\arabic{equation}}

\title{
\hfill\parbox{4cm}{\normalsize IMSC/2006/02/02\\
                               hep-th/0602090}\\        
\vspace{2cm}
Connection between the Loop Variable Formalism and the Old Covariant Formalism
for the Open Bosonic String. 
\author{B. Sathiapalan\\ {\em Institute of Mathematical Sciences}\\
{\em Taramani}\\{\em Chennai, India 600113}\\ bala@imsc.res.in}}           
\maketitle     

\begin{abstract}

The gauge invariant loop variable formalism and old covariant formalism for bosonic open
 string theory are compared in this paper.
It is expected that for the free theory, after gauge fixing, the loop variable
 fields can be mapped to those of the old covariant formalism
in bosonic string theory, level by level. This is verified  explicitly for the
 first two massive levels. 
It is shown that (in the critical dimension) the fields, constraints and 
 gauge transformations can all
be mapped from one to the other. Assuming this continues at all levels 
one can give general arguments that the
 tree S-matrix (integrated correlation functions for on-shell physical
 fields) is the same in both formalisms and therefore they
describe the same physical theory (at tree level). 

\end{abstract}

\newpage

\section{Introduction}

A manifestly background independent formalism  would
 be a big step
towards obtaining a deeper understanding of string theory. In particular
issues such as the space time symmetry principle underlying string theory 
and a fundamental role of strings in the structure of space time might be
elucidated if such a formalism were available. Coventionally,
 most of our understanding of string theory
is based on the world sheet theory.  One can derive, mathematically, 
 some symmetry transformations of the space-time fields
of string theory starting from  world sheet properties such as
 reparametrization invariance
or BRST invariance. On the other hand we know that the low energy 
effective action is the Einstein action
for gravity (or possibly it's supersymmetric generalizations).
 One expects therefore that the sought after symmetry principle would be some
generalization of  general
coordinate invariance. We do not have
 an understanding of this today.

The loop variable (LV)  formalism \cite{BSLV, BSREV} 
incorporates gauge invariance 
without relying on world sheet properties.  This is promising from the point
of view of making  background independence manifest. 
 In fact it was  shown
 recently that
within this approach  one can put open strings in a curved space-time
 background. Thus one can
obtain gauge invariant and general covariant equations of motion for
 massive higher spin
fields  in arbitrarily curved space-times. 
Another approach to background independence is described in \cite{Wi,LW}.

Another advantage of this formalism is that
 the gauge transformations
have a simple form of space time scale transformations.  This is
a step in the direction of understanding the space-time symmetry 
principle underlying
string theory.  Some speculations on this are contained in \cite{BSLV}

There are also some intriguing connections with M-theory: the loop variable
formalism is more conveniently written in one higher dimension, and also the
interacting theory seems best described by a free 'band'' .

What has not been done so far is to obtain a precise map from 
this formalism to other
formalisms of string theory.  The field content of this theory is the same as
 that of BRST string field theory \cite{SZ,WS,Wi2}.
Thus one expects that there should be a map at each level
 between the fields (and their gauge transformations) of the two formalisms.
Alternatively, after gauge fixing one can compare LV formalism with the old
 covariant formalism.
This is what is done in this paper.
In the critical dimension the physical state constraints guarantee 
the absence of unphysical states, and therefore unitarity,
in the old covariant (OC) formalism. One expects that at every level
 one can gauge
fix the loop variable fields and map them to the old covariant
 formalism fields. This is because at the free level the field content uniquely
determines the equations and gauge transformations (modulo field redefinitions).
In this paper the existence of such a map is verified 
by explicit construction for the first two massive 
levels. This map
is  between the fields and gauge transformations of the two
 formalisms.
 The gauge transformations
in the loop variable formalism take the simple form 
$ k(s) \rightarrow k(s) \la (s)$.
In the OC formalism it is generated by $L_{-n}$. We show that 
while the LV formalism is
gauge invariant in any dimension, only in the critical dimension do
the gauge transformations agree with those of the OC formalism. 
Similarly the constraints in the OC and LV formalism can also be 
mapped to each other. 
Assuming this agreement works at higher levels as well one, can argue
 that by gauge fixing the LV
fields, and using the physical state constraints, one obtains the same 
vertex operator as in the OC formalism
and therefore the integrated correlation functions match.  
Thus the S-matrices should agree and thus they describe the same theory.

The critical dimension, D=26, enters in a crucial way for the agreement. 
Away from D=26, the LV formalism is still gauge invariant. It is not clear
whether it can be related to some non critical string theory.

This paper is organized as follows. In Section 2 we describe the
 level 2 and 3 vertex operators in the OC formalism and also list the 
constraints and gauge transformations. This is a review of well known results
(see for e.g. \cite{GSW,Poly}).
 In Section 3 we repeat this for the
LV formalism. In Section 4 we discuss the gauge fixing and also 
give the map between the  fields and constraints in the two formalisms. In Section 5
we discuss the interacting theory. Section 6 contains some conclusions.

\section{Old Covariant Formalism}

In this section we will discuss the physical states and gauge transformations
in the OC formalism for level 2 and level 3. The physical state constraints 
are given by the action of $L_{+n}, n\geq 0$ and gauge transformations by 
the action of $L_{-n}, n > 0$. In \cite{BSVG} a closed form expression is given
for the following:
\be       \label{OC} 
\eln \gvk |0 \rangle 
\ee
where $\tY_n = {\p _z^n X (z) \over (n-1)!}~, \tilde Y_0 = \tilde Y =X$. We will need it mainly
to linear order in $\la _n$ which can be obtained from:
\be   \label{Vir}
e^{-{1\over 2}\cal Y^T \la  \cal Y }\gvk
\ee
where ${\cal Y}^T = (..,\tY _3 , \tY _2 , \tY _1, -i k_0 , -i k_1 , -2i k _2 , 
-3i k_3,....)$ and $\la $ is a matrix whose elements are given by:
\[ (\la )_{m,n} = \la _{m+n} \] (\ref{Vir}) will be used below.

\subsection{Level 2}

\subsubsection{ Vertex operators}

The level two vertex operators can be obtained from
\be \label{2}
e^{ik_0 .\tY + ik_1 . \tY _1 + ik_2 .\tY _2} |0\rangle =
 e^{ik_0.X}(... -{1\over 2} \kim \kin \p X^\mu \p X^\nu +
 i \ktm \pp X^\mu  +...) |0\rangle
\ee

\subsubsection{ Action of $L_{\pm n}$}

       Using (\ref{Vir}) we get
\[
exp ~~ [\la _0 ( {k_0^2\over 2} + ik_1 \tY_1 + 2ik_2 \tY _2) + 
\la _{-1}(k_1.k_0
+2ik_2.\tY_1)
\]
\[ + \la _{-2}(2k_2.k_0 + {1\over 2} k_1.k_1) + \la _1 ( ik_1.\tY_2
+ik_0.\tY_1 )
\]
\be   \label{4}
 + \la _2 (-{1\over 2} \tY_1 .\tY_1 + ik_0 \tY_2 )]
e^{ik_0 .\tY + ik_1 . \tY _1 + ik_2 .\tY _2}|0\rangle
\ee

We need to extract terms that have two $k_1$'s or one $k_2$ for $L_{+n}$.
To get gauge transformations $L_{-1} , L_{-2}$ 
we need to extract the level two terms:
We can read off the various terms:

\begin{enumerate}
\item ${\bf \la _0 L_0 :}$ 
\be
\la _0 ~[ {k_0^2 \over 2}~
 (-{1\over 2} \kim \kin \p X^\mu \p X^\nu + i \ktm \pp X^\mu )
 -\kim \kin \p X ^\mu \p X^\nu + 2i \ktm  \pp X ^\mu ]~e^{ik_0.X}|0\rangle
\ee

\item ${\bf \la _{-1}L_1 :}$
\be
\la _{-1} [ k_1.k_0 ~ i\kim \p X ^\mu + 2i \ktm \p X ^\mu ]~e^{ik_0.X}|0\rangle
\ee
 
\item ${\bf \la _{-2} L_2 :}$

\be  \label{7}
\la _2 [ 2k_2.k_0 + {1\over 2} k_1.k_1 ]~e^{ik_0.X}|0\rangle
\ee

\item ${\bf \la _{1} L_{-1} :}$

\be
\la _1 [ i\kim \pp X ^\mu - \kim \kon \p X^\mu \p X^\nu ]~e^{ik_0.X}|0\rangle
\ee
(It is easy to see that the above is just 
$\la _1 L_{-1}i\kim \p X^\mu ~e^{ik_0.X}|0\rangle $)

 \item ${\bf \la _{2} L_{-2} :}$
\be  
\la _2 [ -{1\over 2} \p X . \p X + i \kom \pp X ^\mu ] ~e^{ik_0.X}|0\rangle
\ee
(This is just $ \la _2 L_{-2} e^{ik_0.X}|0\rangle$)
\end{enumerate}

The $L_0 =1$ equation gives the mass shell condition and
$L_1, L_2 V|0\rangle =0$ and  give additional physical state constraints. 
It is also important to observe
  that since $L_n |0\rangle=0, n\geq -1$, the constraints given above
are equivalent to $[L_n, V]=0, n\geq -1$. For the gauge transformations 
$L_{-2} |0\rangle \neq 0$ and it makes a difference whether one defines
gauge transformation on the fields as including the action on $|0\rangle$ as
is being done here, or as only acting on the vertex operators : $[L_{-n},V]$.
In LV formalism one does not include the action on the vacuum. This has
to be accounted for by field redefinitions.

\subsubsection{Liouville Mode}

One can obtain the physical state constraints also by looking at the
Liouville mode dependence. The Liouville mode,, $\rho $ , is related to 
$\la _n$ at linear order by
\be
{d\la \over dt} = \rho
\ee
where $\la (t) = \sum _n \la _n t^{1-n}$ and $\rho (t) = \rho (0)+ 
t \p \rho (0) + {t^2\over 2}\pp \rho (0)+...$. Thus we get
\be  \label{11}
 \la _0 = \rho ,~~ \la _{-1} = {1\over 2} \p \rho ,~~~\la _{-2}={1\over 3!}
\pp \rho ,~~ \la _{-3}= {1\over 4!}\ppp \rho
\ee

This way of looking at the constraints is useful for purposes of
comparison with the LV formalism.

Thus
\[
e^{ik_0 .\tY + ik_1 . \tY _1 + ik_2 .\tY _2}=
:e^{ik_0 .\tY + ik_1 . \tY _1 + ik_2 .\tY _2}: 
\]
\[
e^{{1\over 2}[k_0^2 \langle X X\rangle + 2 k_1.k_0 \langle X \p X \rangle
+ k_1.k_1 \langle \p X \p X \rangle + 2 k_2.k_0 \langle X\pp X \rangle] }
\]
\be   \label{12}
= :e^{ik_0 .\tY + ik_1 . \tY _1 + ik_2 .\tY _2}: 
e^{{1\over 2}[k_0^2 \rho + 2 k_1.k_0 {1\over 2}\p \rho 
+ k_1.k_1 {1\over 6}\pp \rho + 2 k_2.k_0  {1\over 3} \pp \rho] }
\ee
The Liouville mode dependence is obtained using (\ref{11}), (\ref{4}).
This implies $\langle X X\rangle = \rho ,~~ 
\langle X \p X \rangle = {1\over 2}\rho 
,~~ \langle X \pp X \rangle = {1\over 3}\pp \rho ,~~ \langle \p X \p X \rangle =
 {1\over 6}\pp \rho $. These can be derived by other methods also
 \cite{dhoker}.

 In addition to the anomalous dependences, the Liouville mode
also enters at the classical level. The vertex operators on the boundary
involve covariant derivatives $\nabla _x $ where $x$ is the coordinate
 along the boundary of the world sheet.  
 The vertex operators on the boundary should be
  : $\int dx   V$ where $V$ is a  one dimensional vector vertex operator
or $\int dx \sqrt g S$ where $S$ is one dimensional scalar.
Note that $g_{xx} = g$ (in one dimension) and $g^{xx} = {1\over g}$
The simplest vertex operator is thus $\nabla _x X = \p _x X$
(since $X$ is a scalar). Further $\nabla ^x X = g^{xx} \nabla _x X$
and using $\nabla _x T^x = {1\over \sqrt g} \p _x ( \sqrt g T^x)$ we get
$\nabla _x \nabla ^x X = {1\over \sqrt g}\p _x \sqrt g g^{xx} \p _x X$
$= {1\over \sqrt g}\p _x {1\over \sqrt g} \p _x X$
is a scalar. Thus $\sqrt g \nabla _x \nabla ^x X = \p _x
 {1\over \sqrt g}  \p _x X$ is the vertex operator with two derivatives.
One can similarly show that 
$\p _x {1\over \sqrt g} \p _x {1\over \sqrt g}\p _x X$
 is the vertex operator with three derivatives. This pattern continues.

The metric on the boundary is induced by the metric on the bulk:
\be
g_{xx} = 2 {\p z \over \p x}{\p \bar z \over \p x}g_{z\bar z}= g_{z\bar z}
\ee

Thus if in the conformal coordinates $g_{z\bar z}= e^{-2\rho}$, we have
${1\over \sqrt g} = e^\rho$. Thus
\be
\int dx~\p X , ~~\int  dx~e^{\rho}(\pp X + \p \rho \p X ), ~~
\int  dx~e^{2\rho}(\ppp X + 3 \pp X \p \rho
+ \p X \pp \rho ),...
\ee
Or if we remove $\int dx\sqrt g$  we get
\be
e^{\rho}\p X , ~~e^{2\rho}(\pp X + \p \rho \p X ), ~~
e^{3\rho}(\ppp X + 3 \pp X \p \rho+ \p X \pp \rho ),...
\ee 
for the vertex operators. The power of $e^\rho$ now counts the dimension
of the unintegrated vertex operators.
 Inserting this into
(\ref{12}) we get:
\be
= :e^{ik_0 . X + ik_1 .e^\rho \p X + ik_2 .e^{2\rho} (\pp X + \p \rho \p X )}: 
e^{{1\over 2}[k_0^2 \rho + 2 k_1.k_0 {1\over 2}\p \rho 
+ k_1.k_1 {1\over 6}\pp \rho + 2 k_2.k_0  {1\over 3} \pp \rho] }
\ee

This expression gives the complete $\rho$ dependence to linear order. 
One can check that the coefficient of $\la _{-1} = {1\over 2} \p \rho $
is $2i\ktm \p X ^\mu + k_1.k_0 i \kim \p X ^\mu$
and that of $\la _{-2} = {1\over 6} \pp \rho$ is $({1\over 2} k_1.k_1
+ 2 k_2.k_0)$ as required.  

\subsubsection{Space-time Fields}

We can define fields as usual \cite{BSLV,BSREV} by replacing $\kim \kin$ by $\Phi ^{\mu \nu}$
and $\ktm$ by $A^\mu$. The gauge parameters are obtained by replacing
$\la _1 \kim$ by $\epsilon ^\mu$ and $\la _2$ by $\epsilon _2$. Then we have
the following:

{\bf Constraints:}
The mass shell constraint fixes $p^2 + 2=0$. In addition we have,

\begin{enumerate}

\item
\be   \label{17}
p_\nu \Phi ^{\nu \mu} + 2 A^\mu =0
\ee

\item
\be          \label{18}
\Phi ^\nu _\nu + 4 p_\nu A^\nu =0
\ee

\end{enumerate}

{\bf Gauge transformations:}

\[
\delta \Phi ^{\mu \nu} = \eta ^{\mu \nu} \epsilon _2 + 
p^{(\mu}\epsilon ^{\nu )}
\]
\be 
\delta A^\mu = p^\mu \epsilon _2 + \epsilon ^\mu
\ee

Note that the constraints are not invariant under the gauge transformations
unless $\epsilon ^\mu = {3\over 2}p^\mu \eps _2$, along with 
the mass shell condition $p^2 +2=0$ and the critical dimension $D=26$.
 These correspond to
the zero norm states: states that are physical as well as pure gauge.
It is easy to see that this gauge transformation corresponds
to the state $(L_{-2} + {3\over 2} L_{-1}^2)e^{ik_0.X}|0\rangle $.

\subsection{Level 3}

\subsubsection{Vertex Operators}

The vertex operators in th OC formalism at this level can be written down
as follows:
\[
e^{ik_0 .\tY + ik_1 . \tY _1 + ik_2 .\tY _2 + ik_3 \tY _3} |0\rangle =
\]
\be
( ...+ i{k_3^\mu\over 2!}\ppp X^\mu - \ktm \kin \pp X^\mu \p X^\nu -
i {\kim \kin k_1^\rho \over 3!}
 \p X^\mu \p X^\nu \p X^\rho +...)e^{ik_0.X}|0\rangle
\ee 

\subsubsection{ Action of $L_{\pm n}$} 

Using the same equation (\ref{Vir}) one gets:
\[
exp ~~[\la _3 ( ik_0.\tY _3 - \tY_1 . \tY_2) + \la _2 ( ik_1 \tY_3 +
 ik_0 \tY_2 - {\tY_1.\tY_1\over 2} ) +
 \la _1 (i2k_2\tY_3 + ik_1\tY_2 +ik_0\tY_1)
\]
\[
+ \la _0( i3k_3\tY_3 + 2ik_2\tY_2 + ik_1\tY_1 + {k_0^2\over 2})
+ \la _{-1} (i3k_3\tY_2 + i2k_2\tY_1 + k_1.k_0)
\]
\be
+\la _{-2}(i3k_3\tY_1 + 2k_2.k_0 +{k_1.k_1\over 2})
+\la _{-3}( 3k_3.k_0 + 2k_2.k_1)]
e^{ik_0 .\tY + ik_1 . \tY _1 + ik_2 .\tY _2 + ik_3 \tY _3} |0\rangle 
\ee

One can extract as before the action of $L_{+n}$ on the vertex operators
by extracting terms involving $k_3, k_2k_1$ and $k_1k_1k_1$. Similarly 
gauge transformations are obtained by extracting the level three terms.

\begin{enumerate}

\item ${\bf \la _0 L_0:}$

\be
\la _0 (3+ {k_0^2\over 2})[ik_3^\mu \tY_3^\mu  - \ktm \kin \tY_2^\mu \tY_1^\nu
- i \kim \kin k_1^\rho \tY_1^\mu\tY_1^\nu \tY_1^\rho ]
\ee

\item ${\bf \la _{-1} L_1:}$

\be
\la _{-1}[(i3k_3^\mu + i\ktm k_1.k_0 ) \tY_2^\mu -
 (2\ktm \kin + {\kim \kin \over 2}k_1.k_0)\tY _1^\mu \tY_1^\nu ]
\ee

\item ${\bf \la _{-2} L_2:}$

\be
\la _{-2} [i3k_3^\mu \tY_1^\mu + 2k_2.k_0 i\kim \tY_1^\mu +
 {k_1.k_1\over 2}i\kim \tY_1^\mu
\ee

\item ${\bf \la _1 L_{-1}:}$

\be
\la _1 [ i2\ktm \tY_3^\mu - \kim \kin \tY_2^\mu \tY_1^\nu -
 \kom \ktn \tY_1^\mu \tY_2^\nu ]
\ee

\item ${\bf \la _2 L_{-2}:}$

\be
\la _2 [ i\kim \tY_3^\mu - \kom \kin \tY_2^\mu \tY_1^\nu -
i\kim \tY_1^\mu  {\tY_1.\tY_1 \over 2}
\ee

\item ${\bf \la _3 L_{-3}:}$

\be
\la _3 [i\kom \tY_3^\mu - \tY_1 .\tY_2]
\ee

\end{enumerate}
\subsubsection{Liouville Mode}

Exactly as in the level two case one can get the Liouville mode dependences
- both the classical and anomalous terms. 

\[
e^{ik_0 .\tY + ik_1 . \tY _1 + ik_2 .\tY _2+ ik_3 . \tY _3}=
:e^{ik_0 .\tY + ik_1 . \tY _1 + ik_2 .\tY _2 + i k_3. \tY_3 }: 
\]
\[
e^{{1\over 2}[k_0^2 \langle X X\rangle + 2 k_1.k_0 \langle X \p X \rangle
+ k_1.k_1 \langle \p X \p X \rangle + 2 k_2.k_0 \langle X\pp X \rangle
+ 2k_3.k_0 \langle {\ppp X\over 2!}  X\rangle +
 2k_2.k_1 \langle \pp X \p X \rangle ] }
\]
This is the anomalous dependence. Using covariant derivatives gives
the classical part also:
\[
= :e^{ik_0 . X + ik_1 .e^\rho \p X + ik_2 .e^{2\rho} (\pp X + \p \rho \p X )
+ ik_3. e^{3\rho}(\ppp X + 3 \pp X \p \rho+ \p X \pp \rho ) }: 
\]
\be   \label{12}
e^{{1\over 2}[k_0^2 \rho + 2 k_1.k_0 {1\over 2}\p \rho 
+ k_1.k_1 {1\over 6}\pp \rho + 2 k_2.k_0  {1\over 3} \pp \rho
+ 2k_3.k_0 {\ppp \rho \over 8} + 2k_2.k_1 {\ppp \rho \over 12}] }
\ee

We have used $\langle \ppp X X\rangle = {\ppp \rho \over 4}$ and
$\langle \pp X \p X \rangle = {\ppp \rho \over 12}$. Using the connection 
between $\la _n$ and $\p ^n \rho$ one can check that this is the same as
the results of section 2.2.2. In this form it is easier to compare with 
LV formalism.

\subsubsection{Space-time Fields}

We introduce space-time fields as before by replacing $\kim \kin k_1^\rho$
by $\Phi ^{\mu \nu \rho}$, $ \ktm \kin$ by $B^{\mu \nu} + C^{\mu \nu}$
where $B$ is symmetric and $C$ is antisymmetric, and $k_3^\mu$ by $A^\mu$.
For the gauge parameters we let $\la _3$ be $\epsilon _3$, $\la _2 \kim$
be $\eps _{12}^\mu$, $\la _1 \ktm$ be $\eps _{21}^\mu$ and $\la _1 \kim \kin$
be $\eps _{111}^{\mu \nu}$. We then have:

{\bf Constraints:}

The mass shell constraint $L_0=1$ becomes $p^2+4=0$. In addition,

\begin{enumerate}

\item
\be
p_\nu (B^{\mu \nu} + C^{\mu \nu}) + 3A^\mu =0
\ee
\item
\be
{p_\rho \Phi ^{\mu \nu \rho}\over 2} + B^{\mu \nu} =0
\ee

\item
\be
{B^\nu _\nu \over 12} + {p_\nu A^\nu \over 8}=0
\ee

\item
\be
{\Phi ^{\mu \nu}_\nu \over 12} + {A^\mu \over 2} + {p_\nu B^{\nu \mu}\over 3}
=0 
\ee

\end{enumerate}

{\bf Gauge Transformations:}

\[
\delta \Phi ^{\mu \nu \rho} = \eps _{12}^{(\mu }\eta ^{\nu \rho )} + 
p^{(\mu}\eps _{111}^{\nu \rho )}
\]
\[\delta (B^{\mu \nu}+C^{\mu \nu}) = p^\mu \eps _{12}^\nu
+ p^\nu \eps_{21}^\mu + \eps _{111}^{\mu \nu} + \eps _3 \eta ^{\mu \nu}
\]
\be
\delta A^\mu =p^\mu \eps _3  + \eps _{12}^\mu + 2\eps _{21}^\mu
\ee
(A symmetrization has been indicated in the first line, which involves
adding  two other orderings giving three permutations for each term.)

\section{Loop Variable Formalism}

The loop variable formalism is gauge invariant. This requires auxiliary fields
that are obtained by introducing an extra coordinate that we will call
$\theta$ and it's conjugate momentum $q$. This can be thought of as a 
dimensional reduction process in a $D+1$ dimensional theory. The zero mode
$q_0$ becomes the mass of the field. However it is important to note
that the spectrum of string theory requires that it is  $q_0^2$ that has to 
be integers, and not $q_0$ as in usual Kaluza Klein dimensional reduction.

We begin with the following loop variable and define covariantized
vertex operators from it \cite{BSLV,BSREV}:
\be   \label{34}
\lpp = \gvkl
\ee
 where

\[
Y_n = {\p Y \over \p \xn}, ~ n>0,  ~~~~ Y_0 \equiv Y
\]
\be
Y = \sum _{n\ge 0}\aln \tY_n
\ee 
and $\al _n$ are given by $\al (s) = \sum _{n\ge 0}\aln s^{-n}  
 = e^{\sum _{n\ge 0}\xn s^{-n}}$ with $\al _0 = x_0 \equiv 1$. The following
relation is useful: ${\p \aln \over \p \xm} = \al _{n-m}$.

The vertex operator in (\ref{34}) is covariantized in the sense that
there is a well defined action of a gauge transformation on it:
$k(s) \rightarrow k(s) \la (s)$ and equivalently, if we define
$\la (s) = \sum _{n\ge 0} \la _n s^{-n}$ (with $\la _0 \equiv 1$), the gauge 
transformation is: $k_n \rightarrow \kn + \la _p k_{n-p}$.

The equations obtained from this are automatically invariant under this
transformation
because the prescription is to integrate over $\al (s)$. Thus
a multiplication by $\la (s)$ does not affect the final result since it
can be re-absorbed into $\al (s)$. The equations are obtained by demanding
Weyl invariance of the loop variable with Liouville mode dependence.   

The Liouville mode dependence is obtained using 
\be
\langle Y Y \rangle \equiv \SI, ~~\langle \yn Y\rangle =
 {1\over 2}{\p \SI \over \p \xn}, ~~ \langle \yn \ym \rangle = {1\over 2}
({\pp Y \over \p \xn \p \xm} - {\p Y \over \p x_{n+m}}) 
\ee
Here $\SI$ (defined by the first equation)
 is a covariantized version of the Liouville mode
 $\rho$ and is a linear combination of $\rho$ and its
derivatives. As explained in the last section there is also a classical
Liouville mode dependence that one needs to include. In the present
formalism this need not be included. They are obtained by identifying
some of the auxiliary fields that are present with the physical fields.
Thus {\em all} the Lioville mode dependence comes from anomalies.
This will become clear in the examples below. The loop variable
 with it's $\SI$
dependence is thus:
\be
:\gvkl : e^{{1\over 2}[k_0.k_0 \SI + k_n .k_0 ({\p \SI \over \p \xn})
+ {1\over 2}\sum _{n,m >0}\kn .\km 
({\pp \SI \over \p \xn \p \xm} - {\p \SI \over \p x_{n+m}})]}
\ee 

We consider the level two and three operators in turn.
\subsection{Level 2}

\subsubsection{Vertex Operators}

\[
e^{{(k_0^2+q_0^2)\over 2}\SI}[i\ktm \ytm + i q_2 \theta _2 -
 {\kim \kin \over 2} \yim \yin -
{q_1 q_1 \over 2} \theta _1 \theta _1 -  \kim q _1 \yim \theta _1  
\]
\[
+i(\kim \yim + q_1 \theta _1 )(k_1.k_0 + q_1 q_0){1\over 2} {\p \SI \over \p x_1}
+(k_2.k_0 + q_2q_0){1\over 2}{\p \SI \over \p x_2} +
\]
\[ -i {\kim \kin \over 2} q_1 \yim \yin \theta _1 - i {q_1q_1\over 2} \kim \theta _1 \theta _1 \yim \]
\be
 {(k_1.k_1 + q_1q_1)\over 2}{1\over 2}
({\pp \SI \over \p x_1^2} - {\p \SI\over  \p x_2})]e^{ik_0.Y}  
\ee

 Weyl Invariance is independence of $\SI$. The coefficients of
$\SI$ and its derivatives have to be set to zero. There are the constraints.
It will be seen that field redefinitions will make these equivalent to the
constraints of the OC formalism (\ref{17}) and (\ref{18}). This implies that
the classical Liouville mode dependence is included here indirectly
through terms involving $q_n$.

\subsubsection{Space-time Fields, Gauge Transformations and Constraints}

\begin{itemize}
\item{\bf Space-time Fields:}

The fields are obtained by setting $\kim \kin \approx S_{11}^{\mu \nu}$,
 $\ktm \approx S_2^\mu$,
$\kim q_1 q_0 \approx S_{11}^\mu$, $q_1q_1 \approx S_{11}$, 
and $q_2q_0\approx S_2$.
The gauge parameters are $\la _2 \approx  \Lambda _2$, 
$\la _1 \kim \approx \Lambda _{11}^\mu$,
$\la _1 q_1 q_0 \approx \Lambda _{11}$. 

\item{\bf Gauge Transformations:}

\[
\delta S_{11}^{\mu \nu}  = p^{(\mu}\Lambda _{11}^{\nu )},~~ \delta S_2^\mu =
 \Lambda _{11}^\mu + p^\mu \Lambda _2,~~
\delta S_{11}^\mu = \Lambda _{11}^\mu + p^\mu \Lambda _{11},
\]
\be 
\delta S_2 = \Lambda _2 q_0^2 + \Lambda_{11} ,~~ \delta S_{11} =2\Lambda _{11} 
\ee

Now one can make the following identifications:
 $q_1 q_1 \approx q_2 q_0$, $\kim q_1 \approx \ktm q_0$,
 and $\la _1 q_1 \approx \la _2 q_0$.
This gives: $S_{11}^\mu = 2 S_2^\mu$, $S_{11} \approx S_2$ and 
$\Lambda _{11} \approx 2\Lambda _2$ and the gauge transformations are consistent 
with these identifications.

\item{\bf Constraints:}

 The coefficient of $\SI$ gives the usual mass shell condition $p^2+q_0^2 =0$.
Note that the $(mass)^2$ equals  the dimension of the operator, but the $\SI$
dependence representing this (and also all other $\SI$ dependences) 
comes from an anomaly rather than from the 
classical dependence as in the OC formalism.
\begin{enumerate}
\item Coefficient of ${\pp \SI \over \p x_1 ^2}$
\be  \label{c1}
k_1 . k_1 + q_1 q_1 = 0 ~~\Rightarrow   S_{11 \mu}^\mu + S_2 =0
\ee

\item  Coefficient of ${\p \SI \over \p x_2}$
\be    \label{c2}
k_2.k_0 + q_2 q_0 =0 ~~ \Rightarrow p_\mu S_2^\mu + S_2 =0
\ee

\item  Coefficient of ${\p \SI \over \p x_1 } \yim$

\be  \label{c3}
(k_1.k_0 + q_1 q_0)\kim =0 ~~\Rightarrow p_\nu S_{11}^{\mu \nu} + 2 S_2^\mu =0
\ee

\end{enumerate}

The constraint proportional to $ \theta _1$ is seen to be
 a linear combination of the above.
 The equations of motion
are obtained by setting the variational derivative of 
$\SI$ equal to zero, and are gauge invariant.

\end{itemize}
\subsection{Level 3}

\subsubsection{Vertex Operator}
The complete Level 3 gauge covariantized vertex operator is:

\[
e^{{(k_0^2+q_0^2)\over 2}\SI}[ik_3^\mu Y_3^\mu + i q_3 \theta _3 -
 \ktm \kin  \ytm \yin -
\]
\[
q_1 q_2 \theta _1 \theta _2 -  \kim q_2  \yim \theta _2 -
 \ktm q_1 \ytm \theta _1 -i {\kim \kin k_1^{\rho}\over 3!} 
\yim \yin Y _1^\rho -i{(q_1)^3\over 3!}(\theta _1)^3
\]
\[
+i(\ktm \ytm + q_2 \theta _2 )(k_1.k_0 + q_1 q_0){1\over 2} {\p \SI \over \p x_1}
\]
\[
+i(\kim \yim + q_1 \theta _1 )[(k_2.k_0 + q_2 q_0) {1\over 2} {\p \SI \over \p x_2}
 + {1\over 2}(k_1.k_1 + q_1 q_1){1\over 2}({\pp \SI \over \p x_1 \p x_1} -
 {\p \SI\over  \p x_2})]
\]
\be
+(k_3.k_0 + q_3q_0){1\over 2}{\p \SI \over \p x_3} + (k_2.k_1 + q_2q_1)
{1\over 2}({\pp \SI \over \p x_1 \p x_2} - {\p \SI\over  \p x_3})]e^{ik_0.Y}  
\ee

\subsubsection{Space-time Fields, Gauge Transformations and Constraints}
\begin{itemize}
\item{\bf Space-time Fields}
\[
\kim \kin k_1^\rho \approx S_{111}^{\mu \nu \rho},~~ \kim \kin q_1 q_0 
\approx S_{111}^{\mu \nu},
 ~~ \ktm \kin \approx S_{21}^{\mu \nu}
\]
\[
\kim q_1 q_1 \approx S_{111}^\mu ,~~ \kim q_2 q_0 \approx S_{12}^{\mu},
 ~~\ktm q_1q_0 \approx S_{21}^\mu, ~~ k_3^\mu \approx S_3^\mu
\]
\be
q_3 q_0 \approx S_3,~~ q_2 q_1 \approx S_{21},~~ q_1q_1q_1q_0 \approx S_{111}
\ee
\item{\bf Gauge Parameters}
\[
 ~~\la _1 \kim \kin \approx \Lambda _{111}^{\mu \nu},~~ \la _2 \kim \approx 
\Lambda _{12}^\mu, 
\]
\[
\la _1 \ktm \approx \Lambda _{21}^\mu,~~ \la _1\kim q_1q_0 \approx \Lambda _{111}^\mu
\]
\be
 \la _3 \approx \Lambda _3,~~\la _2 q_1q_0 \approx \Lambda _{12} ,
~~\la _1 q_1q_1 \approx \Lambda _{111},~~\la _1 q_2 q_0 \approx \Lambda _{21}
\ee

\item{\bf Gauge Transformations}
\br
\delta S_{111}^{\mu \nu \rho}       & ~= ~&
 p^{(\mu}\Lambda_{111}^{\nu \rho )}          \nonumber \\
\delta S_{111}^{\mu \nu}       & ~= ~& 
4 \Lambda _{111}^{\mu \nu} + p^{(\mu} \Lambda _{111}^{\nu )}      \nonumber \\      
\delta S_{21}^{\mu \nu}       & ~= ~&  
 \Lambda _{111}^{\mu \nu} + p^\mu \Lambda _{12}^\nu + p^\nu \Lambda _{21}^\mu   
     \nonumber \\
\delta S_{111}^\mu       & ~= ~& 
2\Lambda _{111}^\mu + p^\mu \Lambda _{111}            \nonumber \\
\delta S_{12}^\mu       & ~= ~&
 \Lambda _{111}^\mu + 4 \Lambda _{12}^\mu + p^\mu \Lambda _{21}     
      \nonumber \\  
\delta S_{21}^\mu       & ~= ~& 
\Lambda _{111}^\mu + 4 \Lambda _{21}^\mu + p^\mu \Lambda _{12}       
    \nonumber \\
\delta S_3 ^\mu       & ~= ~& 
  \Lambda _{21}^\mu + \Lambda _{12}^\mu + p^\mu \Lambda _3    
     \nonumber \\
\delta S_{111}       & ~= ~& 12 \Lambda _{111}           \nonumber \\
\delta S_{21}       & ~= ~&  \Lambda _{21} + \Lambda _{111} + \Lambda _{12}   
       \nonumber \\
\delta S_{3}       & ~= ~&   \Lambda _{21}+ \Lambda _{12} + 4 \Lambda _3        
\er

\item{\bf Constraints}

\begin{enumerate}
\item
\be    
k_3.k_0 + q_3q_0 =0 ~~\Rightarrow p_\nu S_3^\nu + S_3 =0
\ee
\item
\be
k_2.k_1 + q_2 q_1 =0 ~~\Rightarrow S_{21\mu}^\mu + S_{21}=0 
\ee
\item
\be
(k_1.k_1 + q_1 q_1 )\kim =0~~\Rightarrow S_{111\nu}^{\nu \mu} + S_{111}^\mu =0 
\ee
\item
\be
(k_2.k_0 + q_2 q_0)\kim =0 ~~ \Rightarrow p_\nu S_{21}^{\nu \mu} +
 S_{12}^\mu =0 
\ee
\item
\be
(k_1.k_0 + q_1q_0)\kim \kin =0~
~\Rightarrow p_\rho S_{111}^{\mu \nu \rho} + S_{111}^{\mu \nu} =0
\ee
\item
\be 
(k_1.k_0 +q_1.q_0)\ktm =0~~\Rightarrow p_\nu S_{21}^{\mu \nu}+ S_{21}^\mu =0
\ee
\item
\be
(k_2.k_0 + q_2 q_0) q_1 q_0 =0 ~~\Rightarrow p_\nu S_{21}^\nu + 4 S_{21}=0
\ee
\end{enumerate} 
Note that the last constraint comes from ${\p \SI \over \p x_2}\theta _1$.

\end{itemize}

\section{Mapping from OC Formalism to LV Formalism}

\subsection{Level 2}

\subsubsection{Mapping of fields}

The mapping is given by:

\[
\Phi ^{\mu \nu } = S_{11}^{\mu \nu} + (\eta ^{\mu \nu} + 
{5\over 2}p^\mu p^\nu)S_2
\]
\be   \label{53}
A^\mu = S_2 ^\mu + 2 p^\mu S_2
\ee

\subsubsection{Mapping Gauge Transformations}

If one makes a LV gauge transformation with parameters $\Lambda _{11}^\mu$
 and $\Lambda _2$ one obtains:

\[
\delta \Phi ^{\mu \nu} = p^{(\mu}\Lambda _{11}^{\nu )} + (\eta ^{\mu \nu} +
 {5\over 2} p^\mu p^\nu )4\Lambda _2
\]
\be         \label{55}
\delta A^\mu = \Lambda _{11}^\mu + 9 ~p^\mu \Lambda _2
\ee

The relative values of the different terms in (\ref{53}),
and therefore in the gauge
transformation (\ref{55}), are fixed by requiring 
that the gauge transformation be 
generated by some combination of  $L_{-n}$'s. 
One can check that (\ref{55}) corresponds to a gauge transformation by
 $4\Lambda _2(L_{-2} + {5\over 4} L_{-1}^2) + \Lambda _{11}^\mu L_{-1}$. 

\subsubsection{Mapping Constraints}

  Now consider the constraint
(\ref{18}). We see that 
\be   \label{45}
p_\nu \Phi ^{\nu \mu} + 2A^\mu = p_\nu S_{11}^{\mu \nu} + 2 S_2^\mu +
 (5-{5\over 2}q_0^2)p^\mu S_2
\ee
Only for $q_0^2=2$ does it become the LV constraint
 $p_\nu S_{11}^{\mu \nu} + 2 S_2^\mu$.
Furthermore

\be   \label{46} 
4p.A + \Phi ^\mu _\mu = 4p.S_2 + S_{11\mu}^\mu  + [(D-{5\over 2}q_0^2) -8q_0^2]S_2
\ee
This should equal
\be      \label{47}
4(p.S_2 + S_2) + S_{11\mu}^\mu + S_2 
\ee
This fixes $D=26$ (using $q_0^2=2$). Thus we see that the while the LV
 equations are gauge invariant
in any dimension, when we require equivalence with OC formalism the critical 
dimension is picked out. 
\subsubsection{Equivalence of OC and LV Formalisms}

Now we can see that the LV formalism is equivalent to the OC formalism: 
Start with a vertex operator in the OC formalism with fields that obey
 (\ref{45}, \ref{46}). 
This implies
that the corresponding LV vertex operator obeys the same constraint 
(\ref{45}). Similarly
(\ref{46}) implies (\ref{47}). This is the sum of two constraints 
of the LV formalism. 
Since the LV formalism is gauge invariant one can choose a gauge
(using invariance under $\Lambda _2$ transformations), where
$S_2$ to equal $-p.S_2$. This implies (by the constraint) that $S_{11\mu}^\mu
+ S_2 =0$. 
 Thus if the fields obey the physical state constraints of the OC formalism,
 then using the gauge invariance, we see that the LV constraints are also
satisfied. In the reverse direction it is easier because we just have to take
a linear combination of two LV constraints (\ref{c1}-\ref{c3})
  to get an OC constraint.  

We can go further in analyzing the constraints. After obtaining $p.S_2+ S_2=0$
using a $\Lambda _2$ transformations, there is a further invariance involving
both $\Lambda _{11}^\mu $ and $\Lambda _2$ with 
$p.\Lambda _{11} + q_0^2\Lambda _2 =0$. This transformation preserves all the
 constraints. (We also have to use the mass shell condition $p^2+q_0^2=0$.) 
Using this invariance we can set $S_2=0$ while preserving 
$p.S_2 +S_2=0=S_{11\mu}^\mu + S_2$. This then implies that 
$p.S_2=S_{11\mu}^\mu =0$. A very similar analysis done on the OC side
using the constraint $4p.A + \Phi ^\mu _\mu =0$ and the gauge transformations
with $\eps ^\mu + {3\over 2 }p^\mu \eps$ that preserves the constraint
(provided $D=26,~~q_0^2=2$):
we can use it to set $p.A$ to zero and so $\Phi ^\mu _\mu=0$.  
Thus on both sides we have a transverse vector and a traceless tensor
obeying $p_\nu \Phi ^{\nu \mu} + 2 A^\mu =0$.

 There are also
terms involving $\theta _n$ on the LV side. But if we focus on the equations
of motion involving only vertex operators with $\yn$, then the two systems
are identical. Since the physical states of the string are conjugate
to these vertex operators (that have only $\yn$), this is all we need to 
describe the physics of string theory. 	

Finally we can use transverse gauge transformations involving $\eps ^\mu$
(i.e with $p.\eps =0$) to gauge away the transverse vector $A^\mu$ 
(and the same thing can be done on the LV side to gauge away $S_2^\mu$). This
leaves a tensor $\Phi ^{\mu \nu}$ which is transverse -
 $p_\nu \Phi ^{\nu \mu} =0$ - and traceless. This is the right number of
 degrees of freedom for the first massive state of the bosonic open string. 

This concludes the demonstration of the equivalence of the OC formalism and LV
formalism for Level 2 at the free level. In the next subsection we discuss 
the Level 3 system.

\subsection{Level 3}

There are four constraints on the OC side that need to be mapped to the LV side.
This involves mapping the fields as well as gauge transformation parameters. 
$q_0^2 =4$ is obviously required from the mass shell condition. Furthermore
mapping all the constraints is possible only when $D=26$. 
The result is the following:

\subsubsection{Mapping Fields}

\br
S_{111}^{\mu \nu \rho} + S_\beta ^{(\mu} \eta ^{\nu \rho )} & =
 & \Phi ^{\mu \nu \rho} \nonumber \\
{S_{111}^{\mu \nu}\over 4} + \eta ^{\mu \nu} S_{\al} -
{1\over 4}p^{(\mu } S_\beta ^{\nu )}& = & B^{\mu \nu} \nonumber \\
S_{21}^{\mu \nu} + \eta ^{\mu \nu} S_{\al}& = & B^{\mu \nu} +
 C^{\mu \nu} \nonumber \\
S_{21}^\mu - p^\mu S_\al& = & 3 A^\mu 
\er
Here $S_\beta ^\mu$ is defined by the requirement
 $\delta S_\beta ^\mu = \Lambda _{12}^\mu$ and $S_\al$ is defined
by $\delta S_\al = \eps ^3$.
Also
\br       
S_{111}^\mu &=& -{1\over 2} S_{12}^\mu - {3\over 2}S_{21}^\mu + 
14 S_3^\mu - 14 p^\mu S_\beta + 2p^\mu S_\al \nonumber \\
S_\beta ^\mu &=& {1\over 2}[{S_{12}^\mu - S_{21}^\mu\over 4} + 
S_3^\mu - p^\mu S_\beta ] \nonumber \\
S_\beta &=& {S_3 \over 4} - 2S_\al \nonumber \\
S_{21} &=& {28\over 3}S_\al 
\er
The gauge parameters obey:
\br
\Lambda _{12} & = & 4\eps ^3 \nonumber \\
\Lambda _{111}^{\mu \nu} & = & \eps_{111}^{\mu \nu} \nonumber \\
\Lambda _{12}^\mu & = & \eps _{12}^\mu \nonumber \\
\Lambda _{21}^\mu & = & \eps _{21}^\mu \nonumber \\
\Lambda _{111}^\mu & = & 3\eps _{12}^\mu + 2\eps _{21}^\mu \nonumber \\
\Lambda _{111} & = & {4\over 3}\Lambda _{12} - \Lambda _{21} \nonumber \\
\Lambda _{21} & = & {7\over 18} \Lambda _{12}
\er

Thus after imposing these relations we have one scalar parameter
 and two vector parameters
and one tensor parameters. This is just enough to compare
 with the OC formalism. Let us see how the states of the
 bosonic string are described:
One can truncate the theory by setting some
gauge invariant combinations of fields to zero. 
The gauge invariant theory 
has two vectors, which can be taken to be $S_3^\mu, S_{12}^\mu$. 
One can truncate the theory so that the other vector
$S_{21}^\mu$ can be expressed in terms of these two,
 and we already have a relation
for $S_{111}^\mu$ in terms of these three in the above equations. 
The truncation preserves the gauge invariance.
These two remaining vectors, $S_3^\mu , S_{12}^\mu$ can further 
be {\em gauged } to zero
to get a gauge fixed theory. 
Also the traceless part of $S_{111}^{\mu \nu}$
can be gauged to zero using $\Lambda _{111}^{\mu \nu}$. 
The constraint $k_2.k_1 + q_2 q_1=0$
implies that $\la _1 (k_1.k_1 + q_1 q_1)=0$. 
Thus the trace of the tensor is related to
the scalar by this constraint. Thus we are left 
with one scalar and one transverse three tensor $S_{111}^{\mu \nu \rho}$.
The three tensor can be decomposed into a traceless part and a trace,
which is a vector. This is the 
right degrees for a covariant description
of a traceless transverse three index tensor \cite{F,SH,WF}. 
There is also one two index anti symmetric
tensor. These are the correct states for the 
third massive level of a bosonic open string.   

Let us see how the constraints are mapped: The constraint
$p_\nu (B^{\mu \nu} + C^{\mu \nu}) + 3A^\mu =0$ becomes (6):
$p_\nu S_{21}^{\mu \nu} + S_{21}^\mu =0$. (If we further choose 
a gauge where the scalar is set to zero - we have already 
truncated the 
theory so that there is only one scalar - then $S_\al=0$ 
and $S_{21}^\mu = 3A^\mu$.)

The second constraint ${p_\rho \Phi ^{\mu \nu \rho}\over 4} +  B^{\mu \nu}=0$
becomes a linear combination of (1) and (5):
 $p_\rho S_{111}^{\mu \nu \rho} + S_{111}^{\mu \nu \rho} +
\eta ^{\mu \nu}{(p.S_3 +S_3)\over 8}=0$. Use a gauge transformation
to choose the scalar field such that $p.S_3 + S_3 =0$. Thus both constraints
(1) and (5) are satisfied.

The constraint ${B^\nu _\nu\over 12} + {p_\nu A^\nu \over 8}=0$ becomes
(2) +(7). One can use $p.\Lambda _{21}$ to gauge (7) to zero, so that
(2) is also satisfied.

The remaining constraint is a linear combination of (3) and (4). One can
use $\Lambda _{12}^\mu $ to set (4) to zero and therefore (3) also.

Thus we have seen that the OC constraints imply the LV constraints.
The reverse is obvious.
Having set one vector to zero and the scalar being constrained to equal
the trace of the tensor, we also see that the field content is the same.

This proves that the vertex operators are exactly the same in both
formalisms.

In the next section we see how this can be generalized to the interacting theory.

\section{Interacting Theory} 

One of the interesting features of the LV formalism 
is that the interacting
theory looks very similar to the free theory. 
In fact if one replaces

\be       \label{sub}
k^\mu (s)  \rightarrow \int _0^R dt~ \bar k^\mu (s,t), ~~\SI \rightarrow \SI + G
\ee  
where $G$ is the regulated Green function,  
one gets the interacting theory.
The LV momentum $\bar k^\mu (s,t)$ is defined in terms of it's modes
 as follows:
\br  \label{bark}
\bar \kim (t) & = & \kim (t) + t \kom (t) \nonumber \\
\bar \ktm (t) & = & \ktm (t) + t \kim (t) + {t^2 \over 2}\kom (t)\nonumber \\
.\nonumber \\
.\nonumber \\
.\nonumber \\
\bar k_n^\mu (t) & = & \sum _0^n k_m^\mu (t) D_m^n t^{n-m} \nonumber \\ 
\bar k_n ^\mu (t)& = & k_n^\mu (t) + (n-1) t k_{n-1}^\mu (t) +...+
 {t^n\over n}\kom (t)
\er
where 
\br
D_m^n & = & ^{n-1}C_{m-1} ,~ n \ge m > 0 \nonumber \\
D_0^n & = & {1\over n} , ~ n\ne 0 \nonumber \\
\er 

This implements the following: In the interacting 
theory there are many
vertex operators defined by their location $t$ on the world sheet boundary.
As an example we have $\kim (t)\p X^\mu (t) e^{ik_0.X(t)}$. We do
a Taylor expansion of these vertex operators about the point $0$. Thus
\[
 e^{ik_0.X(t)+ ik_1.\p X (t)}=
\]
\be
 e^{ik_0.[X(0) + t \p X (0) + {t^2\over 2!} \pp X(0)
+...]+ ik_1 (t)[ \p X (0) + t \pp X(0) + 
{1\over 2!}\ppp X(0)+....]}
\ee
 This Taylor expansion brings all the vertex 
operators to one point and it 
becomes one generalized vertex operator of 
the form $\gvk$ with the more
complicated $\kn$ given in (\ref{bark}). 
This operation is well defined
because we have a world sheet cutoff that regulates the theory and coincident
operators are well defined. 
Once we have done this, we have something that,
mathematically, looks like a single vertex operator. 
In  correlation 
functions this is equivalent to Taylor expanding the Green function $G(0,t)
= G(0,0) + t \p G(0,0) +...$. 
This single vertex operator can then be covariantized using the $\aln$
as described in Section 2 and gauge invariant equations obtained. 
The gauge
transformations are exactly the same as in the free case,
 except for the
replacement of $\kn$ by $\int _0^R dt \bar \kn (t)$ and $\la _n$ by
$\int _0^R dt \la (t)$. There is one technical point: $\bar q_n(t)= q_n(t)$
and one can assume that $G^{\theta \theta}(0,t)=0$ - this is necessary
in order to reproduce string correlation functions \cite{BSREV}. 

Once we have this structure we can implement the results of the previous
section: Represent the product of vertex operators in the OC formalism
as well as in the LV formalism as one vertex operator using the above 
construction.  Take the Level 2  map from OC to LV (\ref{53})
of the free theory and apply it 
 to the Level 2 of this interacting theory.
In order to do this let us first rewrite the mapping in terms
of the loop variable momenta. 
Let $K$ denote the OC variables in the representation (\ref{OC})
 and $k,q$ denote the LV variables. The constraint  (\ref{7}) reads
\[
[ 2k_2.k_0 + {1\over 2} k_1.k_1 ]=0
\]
This has to be mapped to the linear combination of two LV constraints-
\[ 
[k_2.k_0 + q_2 q_0] + {1\over 2}[k_1 . k_1 + q_1 q_1]=0
\]
(We also have the identification $q_1q_1 = q_2 q_0$.)
This is done by means of the mapping (\ref{53}). Expressed in terms
of $k,q,K$ the map reads:
\[
K_1 ^\mu K_1^\nu = \kim \kin + (\eta ^{\mu \nu}+ {5\over 2}\kom \kon )q_2q_0
\]
\be
K_2^\mu = \ktm + 2\kom q_2 q_0
\ee
It is easy to generalize this to the interacting case using the method
described above:
\[
\int _0^1 dt \int _0^1 dt'
\bar K_1^\mu (t)\bar K_1 ^\nu (t') = \int _0^1 dt_1 
\int _0^1 dt_2[\bar \kim (t_1) \bar \kin (t_2)
\]
\[
+ (\eta ^{\mu \nu} + {5\over 2}\int _0^1 ~dt_3 \int _0^1 ~dt_4 \kom (t_3)\kon (t_4) )q_2(t_1)q_0(t_2)]
\]
\be
\int _0^1dt \bar K_2^\mu (t) = \int _0^1 dt \bar \ktm (t) + 2\int _0^1~dt_3 \kom (t_3)  \int _0^1
 dt_1 \int _0^1 dt_2q_2 (t_1)q_0(t_2)
\ee
Note that we have set $R=1$.\footnote{What enters in all the equations is the
 dimensionless ratio $R\over a$ where $a$ is the world sheet cutoff. Setting 
$R=1$ is equivalent to rescaling $a$.} As in all such cases of relations 
between loop variable momenta, the mapping to space-time fields
has to be done in a recursive way. In each equation the highest level field
gets defined in terms of lower level fields. Thus the map between
$\Phi ^{\mu \nu}$ and $S_{11}^{\mu \nu}$ is defined once the map between
$A^\mu$ ($\approx K_1^\mu$) and $\kim , q_1$($\approx S_1^\mu, S_1$)
are known from a lower level equation. This becomes one of the inputs  
at the next stage, which is the map between $\Phi ^{\mu \nu \rho}$ and
$S_{111}^{\mu \nu \rho}$, and so on.

Once we have made the map, one can generalize the earlier sequence of arguments
to establish the equivalence of the S-matrices 
modulo the following assumption - {\em we assume that the
equivalence between the free OC formalism and free LV formalism exists at all levels
in the same way as was shown explicitly for Level 2 and 3.}
 This is a very plausible assumption
because the free theory is determined by the field content - which is the same for
LV formalism and BRST string theory. The gauge transformation is also the same 
modulo field redefinitions. Therefore in the critical dimension 
(which is when BRST string theory
is gauge invariant)
 one expects the two theories to be completely equivalent at the free level.
The sequence of arguments for the equivalence of the formalisms in the 
interacting case is: 
\begin{enumerate}
\item We assume that the external fields of the OC string 
obey physical state conditions. Then the
$\rho$ dependence drops out from the vertex operator. The dependence
on the constant part of $\rho$ is the $L_0$ constraint
- this is the generalization of the mass shell condition for the interacting theory.

\item  In the LV formalism we consider the same set of external physical fields
conjugate to the same vertex operators.
Assume that they obey the  same conditions. There are also additional
auxiliary fields and also gauge invariances.
But these OC constraints are equivalent to the constraints on the LV fields 
after using gauge transformations to select a gauge.
Then the $\SI$ dependence drops out. Thus when the fields of
the OC formalism obey physical state conditions, the LV vertex operator
also has no $\SI$ dependence and, 
in some  gauge where the extra fields are gauge fixed to zero,
is identical to the OC vertex operator.
The $L_0$ constraint is thus the same (i.e. gives the same equation) in both formalisms.

\item  
We can apply the above sequence of arguments to the
interactive vertex operators of the two formalisms:
For the interacting case, in the OC formalism, the usual Ward identities
ensure that if the external fields obey physical state conditions
then integrated correlations involving gauge transformations ($L_{-n}$)
on any vertex operator (or set of vertex operators)
is zero. This is equivalent to saying that
the action of 
$L_n$ on the rest of the vertex operators is zero 
(inside an integrated correlation function - the integration is important)
\footnote{ This is the statement that the state
$|\Psi \rangle = V_N \Delta V_{N-1}\Delta V_{N-2}...\Delta |\Phi _M \rangle$ where
$V_i$ are physical vertex operators, $|\Phi _M \rangle $ is a physical state and
$\Delta$ are the propagators,
satisfies $L_n |\Psi \rangle =0, ~n>0$, provided also that $|\Psi\rangle$ satisfies
$(L_0-1)|\Psi \rangle =0$.}.
This is thus the 
condition that the interacting vertex operators defined earlier
in this section (by the substitution
(\ref{sub}) is independent of  $\rho$ . But in that case
so is the interacting LV vertex operator independent of $\SI$
- since the algebraic manipulations relating the two 
are exactly the same in the free
and interacting cases. Thus to summarize: If the external fields
are physical, then the OC interacting vertex operator is $\rho$ independent
and the LV interacting vertex operator is $\SI$ independent.  

\item
The equations of motion are obtained as the coefficients of a particular 
vertex operator.
When the external fields obey the $L_n$ constraints then what remains of 
the equation
of motion is just the $L_0$ condition for the interacting theory.
This we have seen is the same in both formalisms. When we
relax the $L_n$ constraints on external fields and work in a general gauge,
 the equation
of motion in the LV formalism picks up two kinds of terms:

i) terms proportional to the constraints themselves.
This does not affect the physics, because we know 
that for physical external fields these terms are zero, and only physical
fields are required for the S-matrix. So these terms cannot affect the S-matrix 
that is implied by these equations of motion.

ii) total derivatives in $\xn$ - these are generated when we integrate by parts on 
$x_n$. These generate LV gauge transformations.
This continues to be true for the interacting theory.
For the free theory we saw that this is equivalent to the action of some 
linear combinations of
$L_{-n}$'s on vertex 
operators. This must continue to be true for the interacting theory
because the algebra is exactly the same. These terms also  do not affect the 
S-matrix - this follows from the usual Ward identities of conformal field theory.

\item Thus the S-matrix implied by the gauge invariant
 equations is equal to the S-matrix
of the gauge fixed theory with physical external fields. 
This in turn is equal to that of the 
OC formalism.

\section{Conclusions}

This paper is a step in giving a  precise mathematical basis
for the connection between the loop variable formalism for string theory
and some of the other formalisms.
In this paper we have given the details of the correspondence between
 the old covariant formalism and the loop variable
formalism for open bosonic string theory. At the free level the LV
formalism has the same field content as BRST string field theory (in D=26)
and 
is also gauge invariant (in any dimension). The free theory is fully determined
by the field content. Therefore one expects that there is a field
redefinition in D=26, that would take one to the other, and also after gauge 
fixing, to the fields of the old covariant formalism.  We verify this 
expectation for the old covariant formalism explicitly,
 by giving the map between the fields, constraints 
and gauge transformations for Level 2 and 3. (By gauge transformations we mean,
in the old covariant formalism, the action of $L_{-n}$.) 
Such a map exists only in D=26. In other 
dimensions one can match constraints or gauge transformations, but not both. 

In the loop variable approach, the interacting theory looks formally
like a free theory with a different set of generalized loop variables. Therefore
one can use the same techniques to show equivalence with the old covariant formulation
of string theory. It is argued
that the integrated correlation functions are the same in both formalisms
as long as the physical state constraints are obeyed by the external fields.
Thus the S-matrices are the same and they describe the same physical theory
-  at tree level. A  rigorous proof can presumably be made along the lines
above though we have not done so in this paper.

There remain many open questions. For instance one would like to understand the map
to the BRST formalism. It is also not clear what the loop variable formalism describes away from
D=26, in particular one can ask whether it describes a consistent non-critical 
string theory. Perhaps the most pressing question is whether one can use it
in a quantitative way to get non trivial solutions to string theory. The manifest
background independence demonstrated (for gravitational backgrounds) 
in \cite{BSCS} may help in this regard.

\end{enumerate}

\end{document}